\documentclass[twocolumn,showpacs,preprintnumbers,amsmath,amssymb]{revtex4}

\usepackage{graphicx}
\usepackage{dcolumn}
\usepackage{bm}

\begin{document}

\title{Kondo effect in quantum dots coupled to ferromagnetic
leads with noncollinear magnetizations}
\author{R. \'Swirkowicz$^{1}$}
\author{M. Wilczy\'nski$^{1}$}
\author{ J. Barna\'s$^{2}$}
\email{barnas@main.amu.edu.pl} \affiliation{$^1$Faculty of
Physics, Warsaw University of Technology,
ul. Koszykowa 75, 00-662 Warszawa, Poland \\
$^2$Department of Physics, Adam Mickiewicz University, ul.
Umultowska 85, 61-614 Pozna\'n, Poland;  \\ and Institute of
Molecular Physics, Polish Academy of Sciences, \\ ul.
Smoluchowskiego 17, 60-179 Pozna\'n, Poland}

\date{\today}

\begin{abstract}
Non-equilibrium Green's function technique has been used to
calculate spin-dependent electronic transport through a quantum
dot in the Kondo regime. The dot is described by the Anderson
Hamiltonian and is coupled either symmetrically or asymmetrically
to ferromagnetic leads, whose magnetic moments are noncollinear.
It is shown that the splitting of the zero bias Kondo anomaly in
differential conductance decreases monotonically with increasing
angle between magnetizations, and for antiparallel configuration
it vanishes in the symmetrical case while remains finite in the
asymmetrical one.
\end{abstract}


\maketitle

{\it Introduction} Electronic transport through quantum dots
attached to nonmagnetic leads was studied extensively in the past
decade - both theoretically and experimentally. Quantum dots,
particularly those including only a few or few tens of electrons
offer a unique opportunity to study single-electron charging
effects and electron-electron correlations. In particular, the
many-body Kondo phenomenon was predicted \cite{glazman88} and then
observed experimentally at low temperatures
\cite{cronenwett98,gores00}. The phenomenon is understood rather
well
\cite{hershfield91,meir93,yeyati93,wingreen94,kang95,palacios97,scholler00,krawiec02},
although some of its features are still a subject of some
discussion in the literature -- for instance, the magnitude of the
magnetic-field-induced splitting of the Kondo peak in the
differential conductance \cite{costi00,moore00,kogan04}.

It has been shown recently that the Kondo anomaly in electronic
transport through a quantum dot attached symmetrically to
ferromagnetic leads becomes suppressed in the parallel
configuration \cite{martinek03a,lopez03,choi03,martinek03b}, while
in the antiparallel configuration it has features similar to those
in the case of quantum dots attached to nonmagnetic leads. This
behavior has also been observed experimentally \cite{pasupathy04}.
The partial or total suppression of the anomaly is a consequence
of an effective exchange field originating from the coupling of
the dot to ferromagnetic electrodes. The exchange field gives rise
to a nonzero spin-splitting of the equilibrium Kondo peak in the
density-of-states (DOS) in the parallel configuration, whereas no
such a splitting occurs for antiparallel alignment (in symmetrical
situations the overall coupling of the dot to the leads is
independent of the spin orientation for antiparallel alignment).

The splitting and suppression of the Kondo anomaly in electronic
transport through quantum dots attached to ferromagnetic leads was
studied up to now only for collinear magnetic configurations
(parallel and antiparallel). The noncollinear configuration was
considered in Ref. \cite{sergueev02}. However, the approximations
used to derive the appropriate Green functions do not take into
account the exchange-induced splitting of the anomaly. Therefore,
in this letter we address the problem of angular variation of the
zero bias anomaly, i.e., variation of the Kondo peak with the
angle between magnetic moments of the leads. We consider two
situations; a symmetrical one -- like that described above -- and
a nonsymmetrical case corresponding to different strengths and
spin asymmetries of the coupling to the two electrodes.

We show that the the Kondo peak in differential conductance is
suppressed already at small deviations from the antiparallel
configuration. For larger deviations the anomaly varies rather
slowly with the angle between magnetic moments.

 {\it Model and method} We consider a single-level
quantum dot described by a Hamiltonian of the general form
$H=H_{L}+H_{R}+H_{D}+H_{T}$. The terms $H_{\beta}$ describe the
left ($\beta=L$) and right ($\beta=R$) electrodes in the
non-interacting quasi-particle approximation. The term $H_{D}$
stands for the dot Hamiltonian and is of the form,
$H_{D}=\sum_{\sigma}\epsilon_\sigma\,
d_{\sigma}^{+}d_{\sigma}+Ud_{\uparrow}
^{+}d_{\uparrow}d^{+}_{\downarrow}d_{\downarrow}$, where
$\epsilon_\sigma$ is the energy of the dot level (we assume that
the energy level $\epsilon_0$ of the isolated dot is independent
of the spin orientation), and $U$ is the Hubbard-type electron
correlation parameter. The tunneling part, $H_T$, of the
Hamiltonian describes electron transitions between the dot and
electrodes, and is written in the form \cite{rudzinski05}
\begin{equation}
H_T=\sum_{{\mathbf k}\beta}\sum_{ s\sigma}W_{{\mathbf
k}\beta}^{s\sigma}c_{{\mathbf k}\beta s}^+ d_{\sigma} +{\rm h.c.},
\end{equation}
where $c_{{\mathbf k}\beta s}^+$ is the creation operator of an
electron with the wave vector ${\mathbf k}$ and spin $s$ ($s=+$
for majority electrons and $s=-$ for minority ones) in the lead
$\beta$, $d_{\sigma}$ annihilates an electron  on the dot with
spin $\sigma$ ($\sigma =\uparrow ,\downarrow $ along the
quantization axis of the dot), and $W_{{\mathbf
k}\beta}^{s\sigma}$ is the corresponding matrix element. The
matrix $\check{W}_{{\mathbf k}\beta}$ can be written explicitly as
\begin{equation}
   \check{W}_{{\mathbf k}\beta} =
  \left[\begin{array}{cc}  T_{{\mathbf k}\beta +}\cos(\phi_\beta /2)& -T_{{\mathbf k}\beta +}\sin(\phi_\beta /2) \\
    T_{{\mathbf k}\beta -}\sin(\phi_\beta /2) & T_{{\mathbf k}\beta -}\cos(\phi_\beta /2)  \\
\end{array} \right]\,,
\end{equation}
where $\phi_\beta$ is the angle between the local spin
quantization axis in the lead $\beta$ and the quantization axis on
the dot, and $T_{k\beta s}$ are the matrix elements for electron
tunneling from the dot to the spin majority ($s=+$) and spin
minority ($s=-$) electron bands in the lead $\beta$ when
$\phi_\beta =0$. As the quantization axis for the dot we assume
orientation of the exchange field created by the ferromagnetic
electrodes. For a symmetrical and unbiased system one can assume
$\phi_R=-\phi_L=\theta /2$, where $\theta$ is the angle between
magnetic moments of the leads.

Electric current $I$ has been calculated from a formula derived by
Meir et al \cite{meir93} for nonequilibrium situation. This
formula has been extended to noncollinear magnetic configurations.
It includes retarded (advanced) and correlation Green functions of
the dot in the presence of bias voltage applied to the leads
(nonequilibrium situation). To calculate the retarded Green
functions we used the equation of motion method and restricted
calculations to the large $U$ limit. The higher order Green
functions have been decoupled according to the procedure described
by Meir \cite{meir93}. Such an approach gives reasonable results
for temperatures comparable to the Kondo temperature $T_K$.
However, we go beyond this approximation by calculating the Green
functions, renormalized dot level, and occupation numbers
self-consistently. Strictly speaking, the bare (spin-degenerate)
dot level $\epsilon_0$ in the self energy ${\mathbf\Sigma}_1$ was
replaced by the corresponding level spin-split by the effective
exchange field \cite{braun04}
\begin{equation}
   \mathbf{B}_{\rm exch} = (1/g\mu_B)\sum_\beta \mathbf {n}_\beta
   {\rm Re} \int \frac{d\epsilon}{2\pi}(\Gamma^\beta_+-\Gamma^\beta_-)
   \frac{f_\beta (\epsilon )}{\epsilon -\epsilon_0
   -i\hbar /\tau_0},
\end{equation}
where $\mathbf {n}_\beta $ is a unit vector along the magnetic
moment of the $\beta$-th electrode, $\tau_0$ is a relevant
relaxation time \cite{meir93}, $f_\beta$ is the Fermi distribution
functions for the electrode $\beta$, and $\Gamma^\beta_{s}=2\pi
\sum_k\vert T_{k\beta s}\vert^2 \delta (E-\epsilon_{k\beta s})$
for $s=+,-$. In the following $\Gamma^\beta_{s}$ are assumed to be
independent of energy within the electron band and zero otherwise;
$\Gamma^\beta_{s}=\Gamma^\beta_0 (1+sp_\beta )$, where $p_\beta$
is the spin polarization of the lead $\beta$ \cite{martinek03b}.
The renormalization of the dot level plays an important role in
tranport through quantum dots attached to ferromagnetic
electrodes, and leads to the spin splitting and suppression of the
Kondo anomaly in differential conductance
\cite{martinek03a,lopez03,choi03,martinek03b}. The approach based
on the equation of motion with the renormalization procedure
including the exchange field gives results which are fully
consistent with those obtained for collinear configurations by the
perturbational real-time diagrammatic approach \cite{utsumi05}.

For $U\rightarrow\infty$, the equation of motion for the Green
function $\mathbf{G}$ leads to the Dyson equation
$(\mathbf{I}-\mathbf{g}_0
\mathbf{\Sigma})\mathbf{G}=\mathbf{g}_0$, where
$g_{0\sigma\sigma^\prime}=\delta_{\sigma\sigma^\prime}(E-\epsilon_0)^{-1}$
and
$\mathbf{\Sigma}=\mathbf{g}_0^{-1}-\tilde{\mathbf{n}}^{-1}\mathbf{g}_0^{-1}-\tilde{\mathbf{n}}^{-1}
\tilde{\mathbf{\Sigma}}$. Here,
$\tilde{n}_{\sigma\sigma}=1-n_{\bar{\sigma}\bar{\sigma}}$ and
$\tilde{n}_{\sigma\bar{\sigma}}=n_{\bar{\sigma}\sigma}$, where
$\bar{\sigma}\equiv -\sigma$ and $n_{\sigma\sigma^\prime}=\langle
d^+_\sigma d_{\sigma^\prime}\rangle$. The self-energy
$\tilde{\mathbf{\Sigma}}$ is defined as
$\tilde{\mathbf{\Sigma}}=\mathbf{\Sigma}_0+\mathbf{\Sigma}_1$,
where $\Sigma_{0\sigma\sigma^\prime}= -(i/2)\sum_\beta
\Gamma_{\sigma\sigma^\prime}^\beta$ with
$\Gamma^\beta_{\uparrow\uparrow}=\Gamma^\beta_+\cos^2(\phi_\beta
/2)+\Gamma^\beta_-\sin^2(\phi_\beta /2)$,
$\Gamma^\beta_{\downarrow\downarrow}=\Gamma^\beta_+\sin^2(\phi_\beta
/2)+\Gamma^\beta_-\cos^2(\phi_\beta /2)$, and
$\Gamma^\beta_{\sigma\bar{\sigma}}=(1/2)(\Gamma^\beta_+-\Gamma^\beta_-)\sin(\phi_\beta)$.
In turn, $\mathbf{\Sigma}_1$ is given by
\begin{equation}
  \Sigma_{1\sigma\sigma}= \sum_\beta
  \int\frac{d\epsilon}{2\pi}\frac{\Gamma^\beta_{\bar{\sigma}\bar{\sigma}}f_\beta (\epsilon )}
  {E-\epsilon_\sigma +\epsilon_{\bar{\sigma}}-\epsilon +i\hbar
  /\tau_{\bar{\sigma}}},
\end{equation}
\begin{equation}
  \Sigma_{1\sigma\bar{\sigma}}= -\sum_\beta
  \int\frac{d\epsilon}{2\pi}\frac{\Gamma^\beta_{\bar{\sigma}\sigma}f_\beta (\epsilon )}
  {E-\epsilon}.
\end{equation}

The corresponding lesser Green function $G^<$ has been calculated
from the Keldysh equation $G^<=G^r\Sigma^<G^a$ with the self
energy $\Sigma^<$ calculated from the Ng ansatz. Writing
$\mathbf{\Sigma}^r-\mathbf{\Sigma}^a=-i\mathbf {\Gamma}^{\rm eff}$
and
$\mathbf{\Sigma}_0^r-\mathbf{\Sigma}_0^a=-i\mathbf{\Gamma}=-i(\mathbf{\Gamma}^L+\mathbf{\Gamma}^R)$,
the formula for electric current takes the form
\begin{equation}
I=\frac{e}{2\hbar}\int \frac{d\epsilon}{2\pi}{\rm Tr}\{ {\mathbf
{\Gamma}}^L\mathbf{G}^r\mathbf{\tilde{\Gamma}}^R\mathbf{G}^a +
\mathbf{\Gamma}^R\mathbf{G}^r\mathbf{\tilde{\Gamma}}^L\mathbf
{G}^a\} (f_L-f_R),
\end{equation}
where $\mathbf{\tilde\Gamma}^\beta={\mathbf\Gamma}^\beta
\mathbf{\Gamma}^{-1}\mathbf{\Gamma}^{\rm eff}$, and $e$ is the
electron charge.

{\it Numerical results} Let us write $\Gamma^R_0=\alpha\Gamma^L_0
= \alpha\Gamma_0$. Numerical calculations have been performed for
$\epsilon_0=-0.35$, $kT=0.001$, and $\Gamma_0=0.1$ (energy is
measured in the units of $D/50$, where $D$ is the width of
electron band extending from $-D/2$ to $D/2$). Consider first the
symmetrical situation: $\alpha =1$ and $p_L=p_R=p$. In Fig.1 we
show the corresponding DOS for unbiased (a) and biased (b) system
for parallel ($\theta=0$), antiparallel ($\theta=\pi$), and
perpendicular ($\theta=\pi/2$) magnetic configurations. In the
unbiased system, the spin splitting of the Kondo peak in DOS is
maximal in the parallel configuration and decreases with
increasing angle between the magnetic moments, vanishing exactly
for the antiparallel configuration. This splitting is a direct
consequence of the exchange field exerted on the dot by
ferromagnetic electrodes. The largest exchange field occurs in the
parallel configuration, so the corresponding splitting of the
anomaly is also maximal in this configuration. In the antiparallel
configuration the exchange fields from the two electrodes cancel
each other so the splitting vanishes (see Fig.1(a)). A further
consequence of the spin splitting of DOS is a suppression of the
zero bias anomaly in the differential conductance, as will be
discussed below.

In the biased system, the Kondo peak in DOS becomes split due to
different electrochemical potentials of the two electrodes. Such a
splitting exists also in nonmagnetic situations. The exchange
field in parallel and noncollinear configurations introduces
further splitting, as shown in Fig.1(b), similarly as an external
magnetic field does in the case of quantum dots attached to
nonmagnetic leads \cite{costi00}.

Numerical results for the corresponding differential conductance
and tunnel magnetoresistance (TMR) {\it vs} bias voltage are shown
in Fig.2 for several values of the angle $\theta$ between magnetic
moments of the leads, and for two values of the spin polarization
factor $p$. In the antiparallel configuration ($\theta =\pi$)
there is a well defined zero bias anomaly due to the Kondo
correlations. When the angle between magnetic moments decreases,
the conductance peak becomes split and its intensity becomes
suppressed. The splitting as well as the suppression of the
intensity grow with decreasing $\theta$. Finally, in the parallel
configuration only low-intensity peaks occur at nonzero positive
and negative bias. The results are consistent with the limiting
situations of collinear configurations, $\theta =0$ and $\theta
=\pi$, considered in Refs
[\onlinecite{martinek03a,choi03,martinek03b,utsumi05}]. It is
interesting to note a fast decrease of the peak intensity already
at small deviations from the antiparallel configuration, as
follows from Fig.2(a) and (b). Variation of the peak intensity
with $\theta$ is rather small for $\theta <\pi /2$. Comparison of
Fig.2(a) and (b) also shows that the splitting increases with
increasing spin polarization of the leads, while the peak
intensity decreases with increasing $p$.

\begin{figure}[!ht]
\begin{center}
\includegraphics[width=0.25\textwidth]{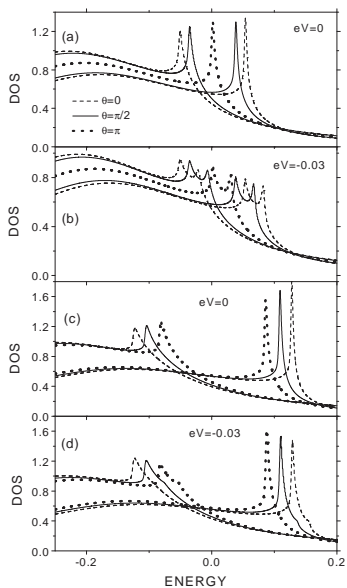}
\end{center}
\caption{Density of states in the Kondo regime in (a,b)
symmetrical ($\alpha =1$, $p=0.2$) and (c,d) asymmetrical ($\alpha
=0.1$, $p_L=0.4$, $p_R=0.8$) systems for three indicated values of
the angle between magnetic moments and for unbiased (a,c) and
biased (b,d) systems. The other parameters as described in the
text} \label{fig1}
\end{figure}

The associated TMR ratio, defined by the system resistance
$R(\theta )$ as $[R(\theta)-R(\theta =0)]/R(\theta =0)$, shows  a
peculiar behavior displayed in Fig.2 (c) and (d) for two values of
$p$. For $\theta =\pi$ the TMR has a dip in the zero bias regime
and is negative. The negative sign is a consequence of the
suppression of the Kondo peak in the parallel configuration, which
makes the system more conducting in the antiparallel configuration
(inverse TMR effect). The dip at zero bias disappears when the
initial configuration departs from the antiparallel one, and
instead of a minimum there is a local maximum of TMR at $V=0$. Two
side local minima of TMR occur then at some nonzero voltages. The
maximum at $V=0$ increases and becomes more flat when the angle
between magnetic moments decreases from $\theta =\pi$ to $\theta
=0$.

\begin{figure}[!ht]
\begin{center}
\includegraphics[width=0.37\textwidth]{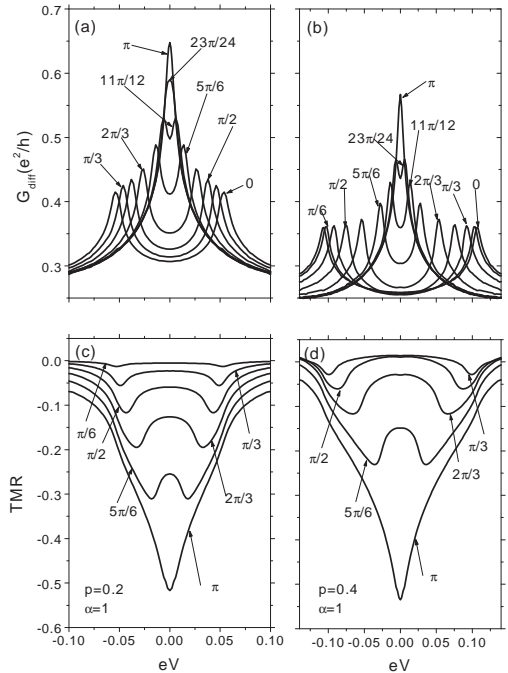}
\end{center}
\caption{Bias dependence of the Kondo anomaly in differential
conductance (a,b) and of the TMR effect (c,d), calculated for
indicated values of the angle $\theta$ between magnetic moments of
the leads and for two different polarization factors $p$. The
other parameters as in Fig.1} \label{fig2}
\end{figure}

In the situation described above the quantum dot was coupled
symmetrically (in the parallel configuration) to the leads. When
the coupling is not symmetrical, then a certain splitting and
suppression of the zero bias anomaly exist also in the
antiparallel configuration \cite{swirkowicz05,utsumi05}. This is
because the exchange fields  from the two leads do not compensate
each other as they did when the dot was coupled symmetrically.
Accordingly, the splitting of the Kondo anomaly may be observed
also when only one of the leads is ferromagnetic while the other
one is nonmagnetic \cite{swirkowicz05}.

\begin{figure}[!ht]
\begin{center}
\includegraphics[width=0.26\textwidth]{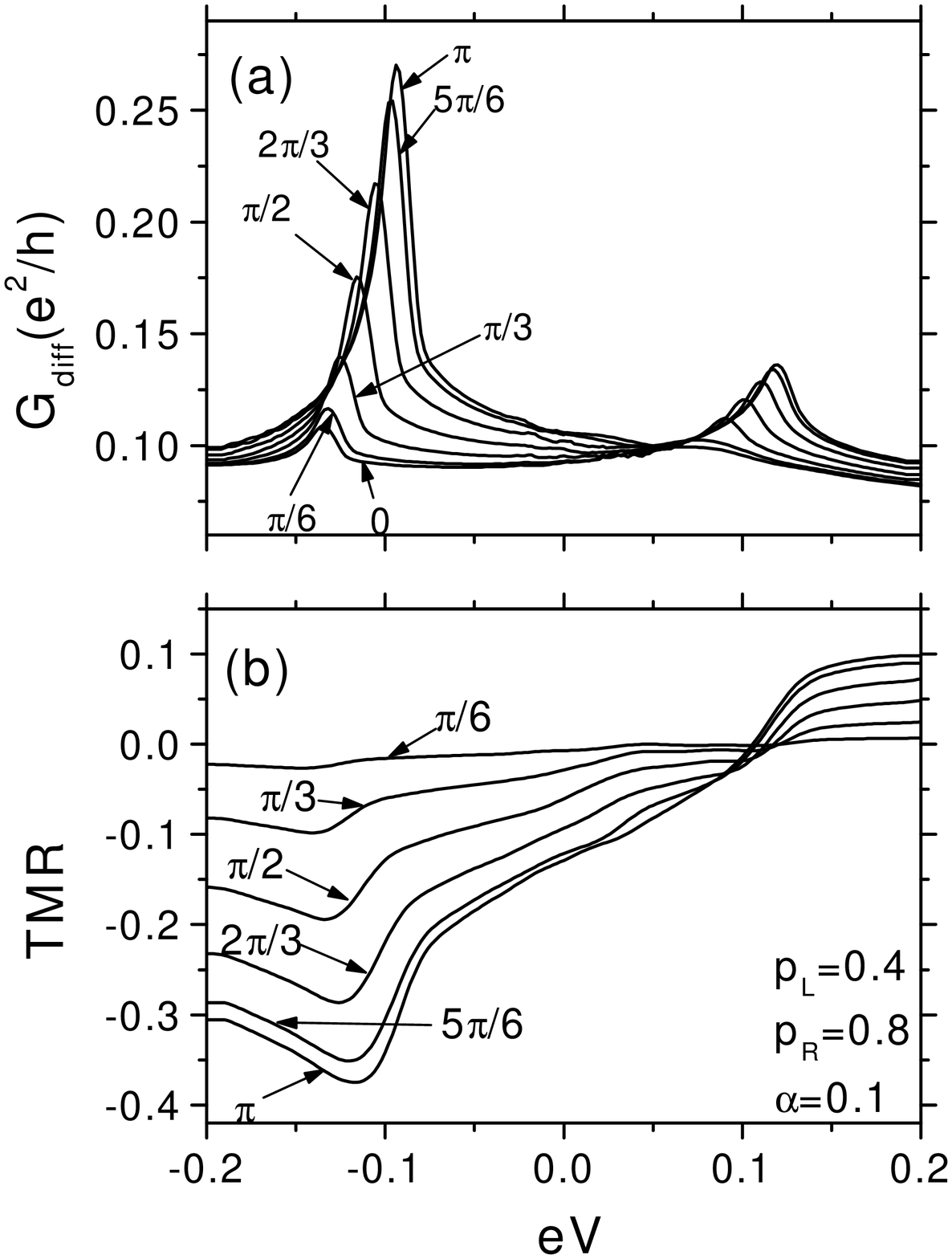}
\end{center}
\caption{Bias dependence of the Kondo anomaly in differential
conductance (a) and TMR (b), calculated for indicated values of
the angle $\theta$ and lead polarization factors.} \label{fig3}
\end{figure}

Let us consider now angular variation of the Kondo anomaly in a
strongly asymmetric system; $\alpha =0.1$, $p_L=0.4$, $p_R=0.8$.
The corresponding DOS is shown in Fig.1(c) and (d) respectively
for the unbiased and biased system, and for parallel, antiparallel
and perpendicular magnetic configurations. At $V=0$ the Kondo peak
in DOS is spin-split in the antiparallel configuration and the
splitting increases when the configuration varies from
antiparallel to parallel. In the biased system the peaks are
additionally split due to different electrochemical potentials of
the two leads, but intensity of one of the two components of the
peak is relatively small and only weakly resolved in Fig.1(d). The
asymmetry in peak intensities is a consequence of the spin
asymmetry in the coupling of the dot to metallic electrodes, and a
difference in spin polarizations, $p_L\ne p_R$.
The differential conductance and TMR are shown in Fig.3 (a) and
(b), respectively. In the parallel configuration the dominant
Kondo anomaly occurs in the spin-up channel and for $eV>0$
(electrons flow from left to right). The corresponding peak at
$eV<0$ is significantly weaker. In the antiparallel configuration,
on the other hand, the Kondo peak in differential conductance
occurs for $eV<0$ only. This behavior can be accounted for by
taking into account behavior of the Kondo peaks in DOS and spin
asymmetry in the coupling to electrodes \cite{swirkowicz05}.

The corresponding TMR is shown in Fig.3(b). It is interesting to
note that TMR is highly asymmetric with respect to the bias
reversal. It becomes positive for $eV$ exceeding a certain
positive value, and negative below this voltage. This is a
consequence of the fact that for positive $eV$ the Kondo peak in
differential conductance is clearly visible in the parallel
configuration, whereas for $eV<0$ the Kondo peak occurs in the
antiparallel configuration. Such a behavior of the conductance and
TMR may be interesting from the point of view of applications in
mesoscopic devices like diodes, for instance.

In summary, we have analyzed the Kondo phenomenon in a quantum dot
attached to ferromagnetic leads with noncollinear magnetizations.
Our results show a monotonic dependence of the  Kondo peak
suppression with the angle between magnetic moments of the leads.
A fast decrease of the Kondo peak takes place already at small
deviations from the antiparallel configuration.

This work was supported by EU trough RTNNANO contract No
MRTN-CT-2003-504574, RTN 'Spintronics' HPRN-CT 2002-00302.

.

\end{document}